# ACOUSTIC SIGNAL BASED NON-CONTACT BALL BEARING FAULT DIAGNOSIS USING ADAPTIVE WAVELET DENOISING

*Wonho Jung, Jaewoong Bae and Yong-Hwa Park*

Department of Mechanical Engineering, Korea Advanced Institute of Science and Technology, Daejeon, 34141, South Korea

## ABSTRACT

This paper presents a non-contact fault diagnostic method for ball bearing using adaptive wavelet denoising, statistical-spectral acoustic features, and one-dimensional (1D) convolutional neural networks (CNN). The health conditions of the ball bearing are monitored by microphone under noisy condition. To eliminate noise, adaptive wavelet denoising method based on kurtosis-entropy (KE) index is proposed. Multiple acoustic features are extracted base on expert knowledge. The 1D ResNet is used to classify the health conditions of the bearings. Case study is presented to examine the proposed method's capability to monitor the condition of ball bearings. The fault diagnosis results were compared with and without the adaptive wavelet denoising. The results show its effectiveness of the proposed fault diagnostic method using acoustic signals.

*Index Terms*— Non-contact fault diagnosis, Wavelet denoising, Acoustic features, Ball bearing, Deep learning

## 1. INTRODUCTION

The reliability of rotating machinery can be attributed to mechanical faults such as bearing wear-out and lubricant deterioration. For consumer electronics products having rotating machinery, the reliability of ball bearing should be evaluated in the qualification testing. However, the qualification test requires a long time. It is a serious obstacle to rapid deployment of a new product. To address this obstacle, the scheme of accelerated life testing can be employed with the health condition monitoring of the ball bearing using vibration sensors. However, vibration sensors have severe data changes depending on the attachment location [1]. These problems make it difficult for non-experts to proceed the accelerated life testing. To solve this problem, a diagnostic method for ball bearing using a non-contact sensor such as a microphone is required in the industrial field. However, signals from microphone are easily affected by external noise, so there are many difficulties in developing a methodology using microphone.

Noise generated from the mechanical or electrical components degrades the fault diagnosis performance. Therefore, a signal preprocessing process for noise reduction is essential for accurate fault diagnosis. There exist many signal processing methods available for ball bearing fault detection, including envelope analysis [2], kurtogram [3], spectral kurtosis (SK) [4], and singular value decomposition (SVD) [5]. However, all these methods have limitations in detecting non-stationary weak fault features. Due to the above issues about detecting non-stationary weak fault features, the wavelet transforms (WT) is considered in many researches [6, 7]. WT is a very effective tool for non-stationary signal processing methods that can detect a weak signal because WT uses basis wavelets similar to the target signal which has discontinuous changes in time domain [8]. WT has more choices on functions to match specific fault symptoms, which is suitable for fault feature extraction. Recently, the noise reduction method using WT based on its characteristics in signal processing has been conducted [9]. However, there is a difference in noise reduction performance depending on basis wavelet selected from various families of wavelet such as Daubechies, Symlet, and Morlet. Therefore, choosing an appropriate basis wavelet is a crucial step for the application of the WT.

This study presents a non-contact diagnostic method for ball bearing that combines adaptive wavelet denoising, domain-statistical-spectral acoustic features, and 1D CNN classifiers. The goal of this study is to develop an acoustic signal-based diagnosis method that can be used to evaluate the durability of ball bearing during product qualification under noisy condition. This paper is organized as follows. In Section 2, the framework of the proposed diagnostic method is described. This section describes two points: (1) how to optimize basis wavelet in terms of denoising, and (2) what is the statistical-spectral acoustic features. In Section 3, a case study using KAIST bearing testbed is presented to evaluate the proposed method. Section 4 concludes this work with future works.

## 2. PROPOSED DIAGNOSTIC METHOD

This section presents a method that combines adaptive wavelet denoising method and statistical-spectral acoustic features for bearings fault detection.

### 2.1. Adaptive Wavelet Denoising

Kurtosis is the normalized version of the fourth central statistical moment of the distributions, it is large for 'spiky' or impulsive signals, because of the considerable weighting given to local spikes by taking the fourth power [10]. Therefore, Kurtosis is especially used to reflect the mechanical incipient fault, because most fault features are sensitive to sharp changes of signals, such as impulses. Thus, the maximum kurtosis of the detailed coefficients is recommended in seeking the optimal wavelet functions. However,

This work was supported by "Human Resources Program in Energy Technology" of the Korea Institute of Energy Technology Evaluation and Planning (KETEP), granted financial resource from the Ministry of Trade, Industry & Energy, Republic of Korea. (No. 20204030200050) the maximum kurtosis of detailed coefficients is not always effective for guiding the adaptive construction of WT. When the ball bearing emerges incipient fault, kurtosis increases obviously. However, kurtosis decreases and remains relatively stable in the later fault period. Therefore, kurtosis index is not effective in detecting periodical impulses.

In information theory, the information entropy of a random variable is the average information provided by each variable and the average uncertainty of the information source [11]. The information entropy can provide useful knowledge on the status of dynamic process. The mechanical fault represents periodic impulsive features that can be detected through the spectrum. At this point, when the fault features dominate the signal, the spectral entropy value decreases, while the normal features dominate signal, the spectral entropy value increases. Spectral entropy can reflect the definite degree of the spectrum.

This paper proposed a new denoising index that combines with kurtosis and spectrum entropy, so-called kurtosis-spectrum entropy (KE) index as following Equation (1). If the signal is well filtered, the kurtosis value is large for the early defect, and the spectrum entropy value is small for the late faulty signal. Also, the KE value of the normal signal is smaller than the faulty signal because the normal signal is not impulsive than the faulty signal. Therefore, the maximum KE of detailed coefficients is used to seek the optimal hyperparameters of wavelet denoising.

$$\mathrm{KE} = \left| \frac{K_P}{E_f} \right| = \left| \frac{\frac{\sum_n^N (x(n)-\mu)^4}{\sigma^4(N-1)}}{\sum_k^K p(s_k)\ln p(s_k)} \right| \quad (1)$$

where $K_P$ is the kurtosis value of the signal; $E_f$ is the entropy value of the spectrum; $x(n)$ is the $n^{th}$ data value of the signals; $\mu$ is the mean value of the signals; $\sigma$ is the standard deviation of the signals; $N$ is the length of the signals; $s_k$ is the $k^{th}$ amplitude of the spectrum; $K$ is the length of the spectrum; and $p$ is the probability density function, respectively.

### 2.2. Statistical-Spectral Acoustic Features and 1D Convolutional Neural Networks

The statistical-spectral acoustic features are composed of time domain, frequency domain, and bearing-related features as shown in Table 1. The time domain features are calculated by characteristic features from time domain signals. It associated with statistical characteristics of the acoustic emission. The frequency domain features are based on the transformed signal in frequency domain. It means energy characteristics of the acoustic emission at the corresponding frequency. The bearing-related features are the energy characteristics excited by defects in bearing elements. The defects in bearings generate impulses with very short duration to the rotating system.

The frequency domain features of bearing defect are extracted in frequency domain. There were two issues when the feature values are extracted. First, the frequency range associated with individual features should be selected carefully. The frequency related to each feature is not fixed but fluctuating as the actual speed of the electric motor changes over time. In order to address this problem, the frequency range related to each feature is determined so that the frequency ranges of two different features do not overlap each other.

Second, the features should be normalized to avoid the numerical instability. It is desirable to scale the magnitudes of different features so that they are in the same order. Several methods can be used to normalize the features. In this study, fourteen features with respect to individual bearing conditions are normalized using the following equation.

$$\overline{X}_{ij} = \frac{X_{ij} - \mu_i}{\sigma_i} \quad (2)$$

where $i$ is the feature number; $j$ is the number associated with the conditions of rolling element bearings; $X_{ij}$ is the $i^{th}$ feature value of the $j^{th}$ condition; and $\mu_i$ and $\sigma_i$ are the mean and standard deviation of the $i^{th}$ feature, respectively.

**Table 1.** Statistical-spectral acoustic features.

| Domain | Equation | Description |
|---|---|---|
| Time | $f_1 = \sum_{n=1}^{N} \frac{x(n)}{N}$ | Mean value of signal |
| | $f_2 = \sum_{n=1}^{N} \frac{(x(n)-f_1)^2}{N}$ | Variance of the signal |
| | $f_3 = \sqrt{\sum_{n=1}^{N} \frac{(x(n))^2}{N}}$ | RMS value (power of the signal) |
| | $f_4 = \max(|x(n)|)$ | Maximum value of the signal |
| | $f_5 = \frac{\sum_{n=1}^{N}(x(n)-f_1)^3}{f_2^3(N-1)}$ | A measure of asymmetry of distribution around its mean (3rd order moment; skewness) |
| | $f_6 = \frac{\sum_{n=1}^{N}(x(n)-f_1)^4}{f_2^2(N-1)}$ | A measure of the spikiness or impulsive features of the signal (4th order moment; kurtosis) |
| | $f_7 = \frac{f_4}{f_3}$ | Ratio of max value to the total energy of the signal (crest factor) |
| | $f_8 = \frac{f_3}{f_1}$ | Ratio of RMS value to the mean of the signal (shape factor) |
| | $f_9 = \frac{f_4}{f_1}$ | Ratio of max value to the mean of the signal (impulse factor) |
| | $f_{10} = \frac{f_4}{\left(\frac{1}{N}\sum_{n=1}^{N}\sqrt{|x(n)|}\right)^2}$ | Ratio of max value to the squared mean value of square roots (clearance factor) |
| Frequency | $f_{11} = \max(s(k)_{BPFO})$ | Maximum value between 1st to 5th order moment of the ball pass frequency of the outer race (BPFO) |
| | $f_{12} = \max(s(k)_{BPFI})$ | Maximum value between 1st to 5th order moment of the ball pass frequency of the inner race (BPFI) |
| | $f_{13} = \sum_{k=1}^{K} \frac{s(k)}{K}$ | Mean value of the amplitudes of frequency response |
| | $f_{14} = \sum_{k=1}^{K} \frac{(s(k)-f_{11})^2}{K}$ | Variance of the amplitudes of the frequency response |

1D CNN architectures derives features from one dimensional vectors of the dataset. The 1D CNN can be used to the analysis of a fixed-length period signal data such as audio signals. The statistical-spectral acoustic features have the fixed-length period signal that is suitable for the application of 1D CNN. In this work, ResNet modules are modified. In modified ResNet, the kernel size is 3 by 1. The ResNet architecture consists of 20 layers. A single 1D convolutional layer is in the first layer. Additional nine modified ResNet modules are added. One fully-connected layer is in the last layer.

## 3. CASE STUDY

### 3.1. Experiment Setup

The KAIST bearing testbed has two bearings connected to the motor rotating shaft. The motor rotates at 1505 rpm, doubled through the gearbox, and the bearings rotate at 3010 rpm. The specification of the bearing is NSK 6205 DDU. Five conditions of bearings are composed of normal, outer race fault (0.3 mm and 1.0 mm), and inner race fault (0.3 mm and 1.0 mm). The faults were artificially applied for each crack size as shown in Fig. 1. Acoustic signals were collected from microphone with sampling rate of 51200 Hz. The acoustic signals were collected every two minutes. The microphone installed near the Bearing A housing of the KAIST bearing testbed. In this paper, PCB 378B02 microphone manufactured by PCB Piezotronics is used.

The modified ResNet was trained using the Adam optimizer with the learning rate between 0.001 and 0.0001. The batch size for training and testing was 32. The training data was composed of one-second signal by overlapping the entire signal after adaptive wavelet denoising. The rate of overlapping is ten percent. Training data are converted to statistical-spectral acoustic features and labeled in five classes including normal, outer race fault (0.3 mm and 1.0 mm), and inner race fault (0.3 mm and 1.0 mm). The training data of 3,840 and the test data of 960 are used.

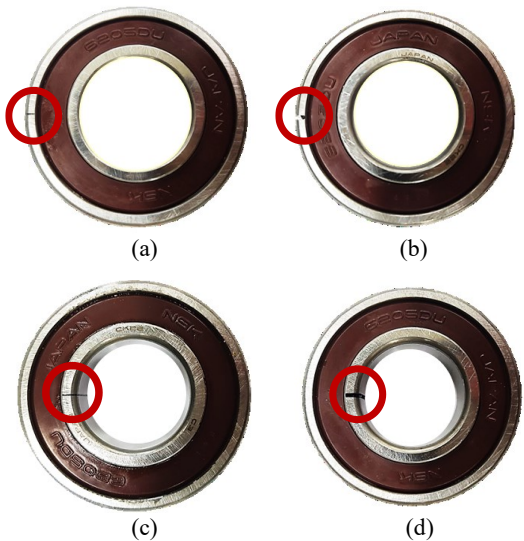

**Fig. 1.** Fault seeding of ball bearing: (a) outer race fault 0.3 mm, (b) outer race fault 1.0 mm, (c) inner race fault 0.3 mm, and (d) inner race fault 1.0 mm.

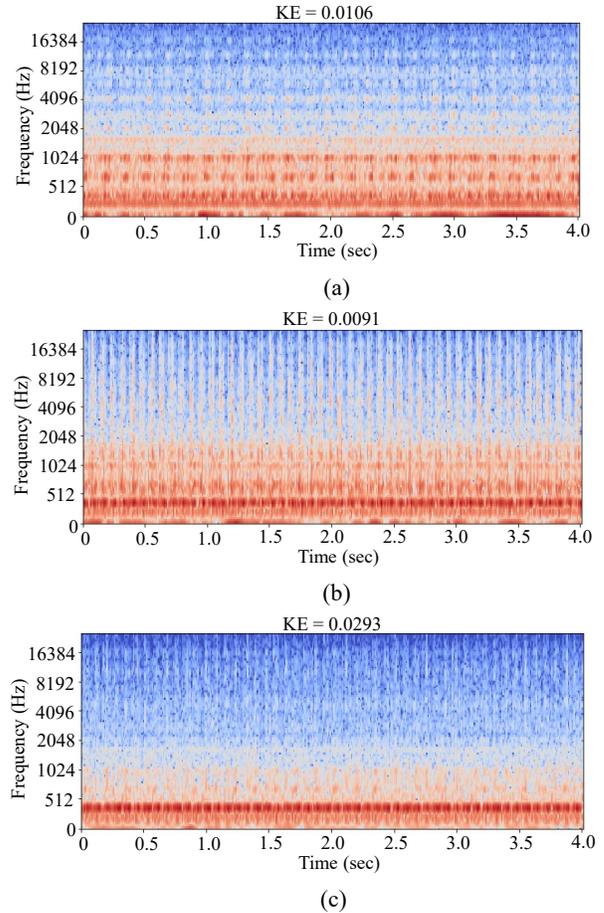

**Fig. 2.** Log Mel-spectrogram of original acoustic signals: (a) normal, (b) outer race fault 1.0 mm, and (c) inner race fault 1.0 mm.

### 3.2. Results and Discussion

The original acoustic signals with log Mel-spectrogram according to each condition of bearing are presented as shown in Fig. 2. The original acoustic signals have a lot of noisy components due to unknown noises from other machinery and external environment, it is difficult to classify the signal with normal or fault label. To tackle this problem, adaptive wavelet denoising method is used. The hyperparameters of wavelet denoising are optimized using the KE index. Optimal hyperparameters are selected to have maximum KE index value among the 270 basis wavelet functions in wavelet families such as Daubechies, Symlet, Coiflet, Morlet, and Biorthogonal wavelet. As a result, 'Biorthogonal3.1' wavelet is selected for normal, 'Coiflet1' wavelet is selected for outer race fault (0.3 mm) and inner race fault (0.3 mm and 1.0 mm). 'Daubechies37' wavelet is selected for outer race fault (1.0 mm). The filtered acoustic signals with log Mel-spectrogram are shown in Fig. 3. In the filtered signals, the characteristics of each outer race defect and inner race defect are prominently displayed. In particular, it can be seen that the defect frequency components for outer ring defects and inner ring defects are prominent and their KE value is higher.

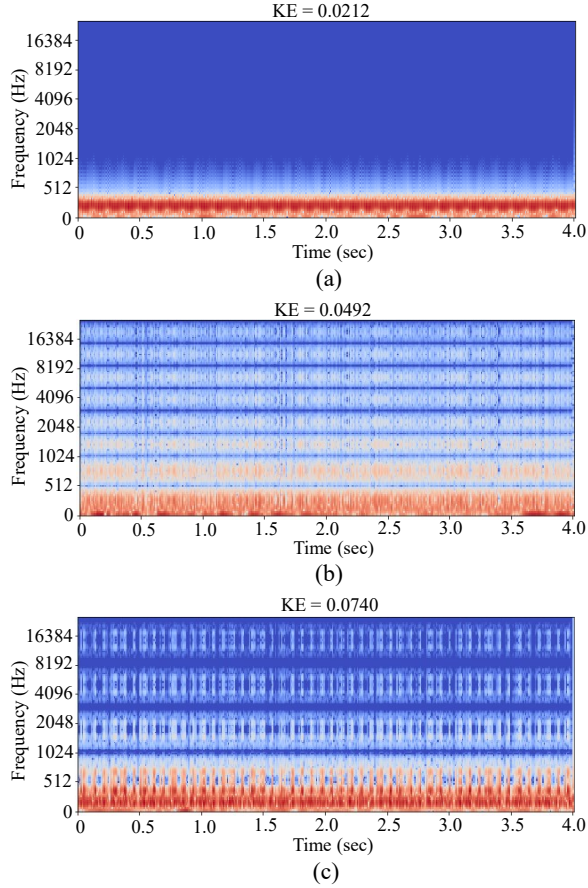

**Fig. 3.** Log Mel-spectrogram of denoised acoustic signals: (a) normal, (b) outer race fault 1.0 mm, and (c) inner race fault 1.0 mm.

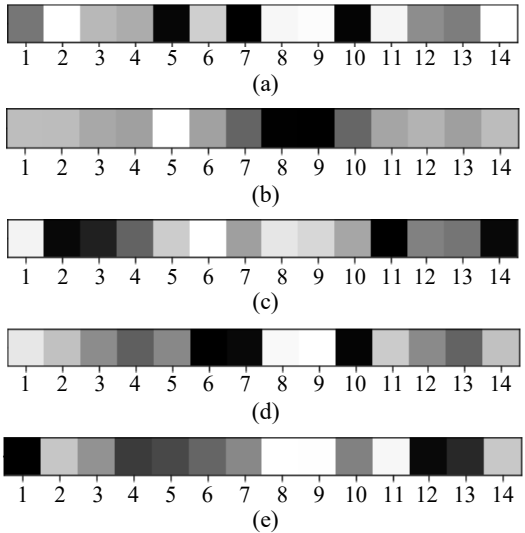

**Fig. 4.** Normalized statistical-spectral acoustic features after adaptive wavelet denoising: (a) normal, (b) outer race fault 0.3 mm, (c) outer race fault 1.0 mm, (d) inner race fault 0.3 mm, and (e) inner race fault 1.0 mm.

**Table 2.** Fault diagnosis accuracy according to data type.

| Data type | Test accuracy |
|---|---|
| Raw acoustic signals without denoising | 52.1 % |
| Raw acoustic signals with denoising | 72.8 % |
| Statistical-spectral acoustic features without denoising | 60.0 % |
| Statistical-spectral acoustic features with denoising | **91.8 %** |

The statistical-spectral acoustic features were visualized as images as shown in Fig. 4. When changing from normal signal to faulty signal, the values of dominant features are changing from the statistical time features to periodic frequency features. The color change of the bearing defect feature such as ball pass frequency of the outer race (BPFO), and ball pass frequency of the inner race (BPFI) can be observed from the 11[th] feature and 12[th] feature in the images. The change is attributed to the increased severity of bearing defects. For example, in outer race fault, the 11[th] feature (BPFO) is more dominant than other features when the severity increase. Also, in inner race fault, the 12[th] feature (BPFI) is more dominant than other features when the severity increase. Therefore, the images in Fig. 3 and Fig. 4 showed that the statistical-spectral acoustic features reflect the characteristics of the motor bearing degradation in life testing by effectively capturing the health condition of ball bearing.

The performance of the modified ResNet-20 was compared to different data types as shown in Table 2. The test accuracy of raw acoustic signals with or without adaptive wavelet denoising method was 72.8 % and 52.1 %, respectively. These results present the adaptive wavelet denoising method works well without losing the defect information. In an actual external noisy condition, normal and faulty signals are easily distorted. Therefore, the consequence of the misclassification when using raw acoustic signals and acoustic features without denoising was not very surprising. In case of using statistical-spectral acoustic features, the test accuracy increases up to 91.8 %. This is attributed that the simple 1D image of the statistical-spectral acoustic features makes easily to classify the health conditions of the bearings. Therefore, the complexity of the ResNet-20 architecture was sufficient to classify the five health conditions of the bearings.

## 5. CONCLUSION

In this work, an acoustic signal based non-contact diagnosis method was proposed to determine the health condition of ball bearing using statistical-spectral acoustic features with adaptive wavelet denoising and 1D CNN. To evaluate the performance of the method, different types of data including raw acoustic signals, acoustic features without denoising method are used. The statistical-spectral acoustic features with proposed adaptive denoising method showed better performance in 1D ResNet-20. This study expected that the non-contact diagnostic approach can be used to evaluate the durability of ball bearing during product qualification using microphone signal under noisy condition. For future studies, a novel feature extraction method for early fault detection based on cyclostationary signals will be studied.